\documentclass[conference]{IEEEtran}
\IEEEoverridecommandlockouts
\usepackage{cite}
\usepackage{amsmath,amssymb,amsfonts}
\usepackage{graphicx}
\usepackage{textcomp}
\usepackage{xcolor}
\usepackage{verbatim}
\usepackage[ruled,vlined]{algorithm2e}

\usepackage{amsthm}

\newcounter{heuristicctr}
\renewcommand{\theheuristicctr}{\arabic{heuristicctr}}
\newsavebox{\heuristicbox}

\newenvironment{heuristic}
  {%
    \refstepcounter{heuristicctr}%
    \begin{center}
    \setlength{\fboxsep}{6pt}%
    \begin{lrbox}{\heuristicbox}%
    \begin{minipage}{0.95\columnwidth}%
    \textbf{Design Rule~\theheuristicctr:}
  }
  {%
    \end{minipage}%
    \end{lrbox}%
    \fbox{\usebox{\heuristicbox}}%
    \end{center}
  }

\newsavebox{\mybox}

\newcommand{\hlsml}{\textsc{hls4ml}\xspace}
\newcommand{\aieml}{\textsc{aie4ml}\xspace}

\def\BibTeX{{\rm B\kern-.05em{\sc i\kern-.025em b}\kern-.08em
    T\kern-.1667em\lower.7ex\hbox{E}\kern-.125emX}}
\begin{document}

\title{Design Rules for Extreme-Edge Scientific Computing on AI Engines}
\author{\IEEEauthorblockN{Zhenghua Ma\textsuperscript{1},
G Abarajithan\textsuperscript{1},
Dimitrios Danopoulos\textsuperscript{2},
Olivia Weng\textsuperscript{1},
Francesco Restuccia\textsuperscript{1},
Ryan Kastner\textsuperscript{1}}
\IEEEauthorblockA{\textit{\textsuperscript{1}University of California San Diego, La Jolla, CA,  USA}}
\IEEEauthorblockA{\textit{\textsuperscript{2}European Organization for Nuclear Research (CERN), Geneva, Switzerland}}
\IEEEauthorblockA{\{zhm007, agnaneswaran\}@ucsd.edu,
dimitrios.danopoulos@cern.ch,
\{oweng, frestuccia, kastner\}@ucsd.edu}
}

\newcommand\zhenghua[1]{\textcolor{green}{\textbf{#1} -Zhenghua}}
\newcommand\francesco[1]{\textcolor{cyan}{\textbf{#1} -Francesco}}
\newcommand\aba[1]{\textcolor{blue}{\textbf{#1} -Aba}}
\newcommand\ryan[1]{\textcolor{red}{\textbf{#1} -Ryan}}

\maketitle
\begin{abstract}
Extreme-edge scientific applications use machine learning models to analyze sensor data and make real-time decisions.
Their stringent latency and throughput requirements demand small batch sizes and require that model weights remain fully on-chip.
Spatial dataflow implementations are common for extreme-edge applications. Spatial dataflow works well for small networks, but it fails to scale to larger models due to inherent resource scaling limitations.
AI Engines on modern FPGA SoCs offer a promising alternative with high compute density and additional on-chip memory. 
However, the architecture, programming model, and performance-scaling behavior of AI Engines differ fundamentally from those of the programmable logic, making direct comparison non-trivial and the benefits of using AI Engines unclear. 
This work addresses \emph{how and when} extreme-edge scientific neural networks should be implemented on AI Engines versus programmable logic.
We provide systematic architectural characterization and micro-benchmarking and introduce a latency-adjusted resource equivalence (LARE) metric that identifies when AI Engine implementations outperform programmable logic designs.
We further propose spatial and API-level dataflow optimizations tailored to low-latency scientific inference.
Finally, we demonstrate the successful deployment of end-to-end neural networks on AI Engines that cannot fit on programmable logic when using the \hlsml toolchain.

\end{abstract}

\section{Introduction}

The scientific community increasingly relies on real-time machine learning (ML) inference to process high-rate sensor data~\cite{duarte2022fastml, frontend-asic, tokamak, reliable}. 
Extreme-edge workloads operate at tens of megahertz and require end-to-end latencies on the order of a few microseconds. 
To meet these constraints, all model weights reside fully on-chip, and inference uses small neural networks (NNs) with a low batch size~\cite{weng2024architectural}. Thus, the scientific community has primarily focused on smaller NNs implemented on FPGA programmable logic (PL) or as an ASIC.

\begin{figure}
    \centering
    \includegraphics[width=0.8\linewidth]{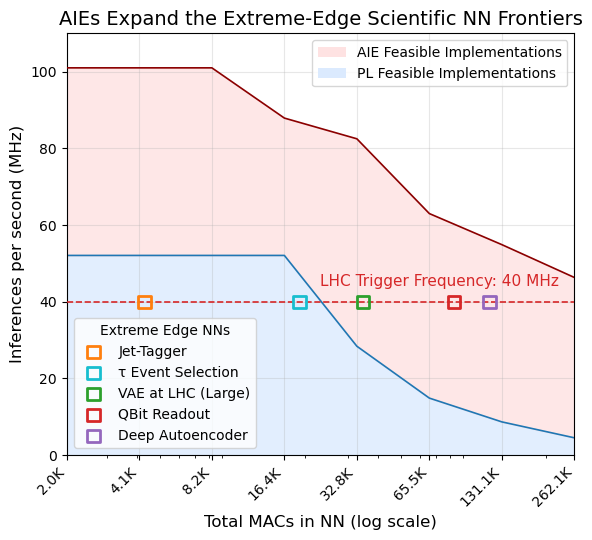}
    \caption{
    Our design rules for AIE allow larger NN implementations to meet and surpass the 40 MHz throughput requirement of the LHC trigger system\cite{govorkova2108autoencoders, cms2016cms}.
    While PL is sufficient for small NNs such as Jet-tagger\cite{jettagger} and $\tau$ Event Selection\cite{misra2025belle2tau}, larger NNs such as VAE\cite{vaelarge}, Qubit Readout\cite{qubit} and the Deep Autoencoder\cite{govorkova2108autoencoders} can only meet the performance requirement on AIEs using our design rules.}
    \label{fig:intro}
\end{figure}

\hlsml~\cite{hls4ml2021} is a popular open-source toolchain for implementing machine learning models on FPGA PL or ASIC. 
\hlsml employs a spatial dataflow architecture where each layer is implemented as a separate datapath, and all weights are stored on-chip. 
%
While this spatial dataflow architecture enables low-latency and high-throughput performance, it is inherently resource-intensive. The spatial mapping scales roughly with the NN's size and depth. 
When implementing larger ML models, PL resources quickly become fully utilized, forcing aggressive reuse of arithmetic units, thereby sharply reducing performance, as shown in Fig.~\ref{fig:intro}. Furthermore, higher reuse factors make HLS optimization problems more challenging, often increasing synthesis tool runtimes to the point of failure. 

The AI Engine (AIE) is a high-performance, 2D array of VLIW vector processors (compute tiles) designed for deterministic, high-throughput execution of digital signal processing and ML workloads~\cite{ahmad2019xilinx}.
AIEs offer a promising alternative to PL for extreme-edge scientific computing, and many in the scientific community are considering AIEs for next-generation applications~\cite{xiotidis2026amd, gonski2026machine}. 
AIEs use a spatial programming model where NN kernels are mapped to the VLIW vector processors and interconnected through data-flow graphs~\cite{amd_ug1076_2025}.
While AIE programming models are being developed~\cite{zhuang2025aries,levental2024mliraie, danopoulos2025aie4ml}, they lack the maturity of PL programming tools.

Although AIEs and PL coexist on the same Versal FPGA SoC\cite{vek280handbook, vckhandbook}, they are fundamentally distinct architectures with unique performance and resource-scaling behaviors, making the benefits of AIE deployment unclear to the extreme-edge scientific community. 
Moreover, existing AIE programming models are primarily optimized for high-throughput, batched General Matrix Multiplication (GEMM) computations ~\cite{charm,charm2.0,maxeva,brown2023exploring,chen2023exploiting} rather than for low-latency, event-driven, weights-on-chip inference required in this domain.
\emph{As a result, the scientific community lacks a systematic methodology to determine when and how to implement extreme-edge NNs on AIEs versus PL. This work aims to bridge this gap.} 

Our key contributions include:

\begin{itemize}
    \item \textbf{Modeling and micro-benchmarking} of workloads between PL and AIEs under realistic resource and performance budgets, capturing the design space of PL parallelism and reuse. We introduce the \emph{Latency-Adjusted Resource Equivalence (LARE)} metric as a decision boundary for determining the tradeoff between AIE and PL implementations. Furthermore, LARE provides insights into when AIEs are underutilized.

    \item \textbf{Tiling and dataflow optimization} for implementing low-latency scientific workloads on AIEs. We derive practical design rules for spatial and API-level tiling from empirical benchmarking. We further quantify architectural bottlenecks including column exhaustion and PL-AIE boundary-crossing overhead, providing guidance for efficient low-latency deployment.

    \item \textbf{Implementation and evaluation} of end-to-end NNs for extreme-edge scientific computing. 
    As shown in Fig.\ref{fig:intro}, with our tiling and dataflow design rules, larger NNs that previously could not meet the trigger frequency on PL can now be implemented successfully on AIEs.  

\end{itemize}

The rest of the paper is organized as follows.
Section~\ref{sec:background} reviews the unique requirements of low-latency scientific applications and motivates our research questions.
Sections~\ref{sec: micro-benchmarking} and \ref{sec:how} present our main contributions, including architectural characterization, micro-benchmarking, and design optimization rules with experimental evaluations.
Section~\ref{sec:nn} demonstrates end-to-end deployment of NNs and the benefits of our design optimizations.
Section~\ref{sec:related} reviews related work. Finally, Section~\ref{sec:conclusion} concludes the paper.

\section{Background and Motivation}
\label{sec:background}

This section describes the unique requirements and state-of-the-art tools in extreme-edge scientific computing. We review spatial dataflow implementations of NNs for high-energy physics applications on FPGAs and their scaling limits for larger NNs. Finally, we present the research challenges and introduce the prospective benefits of leveraging AIEs. 

\subsection{Low-Latency Scientific Edge Computing}
\label{sec:low_latency_computing}
Real-time edge inference for scientific applications imposes strict latency and bandwidth requirements. 
A representative example is the CERN Large Hadron Collider (LHC), built to study fundamental physics by colliding protons at nearly the speed of light.
Its detectors capture hundreds of terabytes of data per second~\cite{cms2008cms, sirunyan2020performance, atlas2020operation, duarte2022fastml}, describing a wide variety of collision events.
However, only a small fraction of these events are scientifically valuable, and storing all detector data is infeasible both in volume and cost. 
Consequently, the system requires filtering near the sensors to retain only the most relevant data.
This requires extreme low-latency and high-throughput processing to keep up with the collisions that occur at a rate of 40 MHz\cite{govorkova2108autoencoders, cms2016cms}. 

FPGA platforms have proven to be effective for such low-latency workloads. 
However, under the above latency and throughput requirements, streaming the weights from off-chip demands bandwidth beyond the capacity of the on-board DRAM interface~\cite{weng2024architectural, zcu102handbook, zcuhandbook, vek280handbook, reliable}. 
Therefore, all weights must be pre-loaded and must remain stationary on-chip throughout inference.

\subsection{Spatial Dataflow Frameworks}
Spatial dataflow frameworks such as \hlsml~\cite{hls4ml2021} and FINN~\cite{finn} streamline the NN deployment on FPGAs. 
In particular, \hlsml was developed to serve the scientific community, providing accessible hardware deployment for users without significant expertise in FPGA or hardware design.
However, spatial dataflow architectures exclusively allocate hardware resources to each layer, leading to resource usage that scales roughly linearly with the NN’s parameter count and depth. 
While designers can tune the reuse factor--that is, how many operations are time-multiplexed onto the same arithmetic units--to trade performance for resource savings, large NNs inevitably require higher reuse factors, which substantially increase latency.
Beyond a certain scale, even aggressive reuse cannot prevent EDA tools from failing to meet resource constraints. 
This challenge considerably limits the size of the NNs that can be deployed on PL.
\subsection{AIE Benefits and Research Challenges}

The AIE architecture offers a promising solution to the scalability challenge. For instance, the Versal VEK280 board~\cite{vek280handbook} integrates 304 compute tiles of AIE-ML architecture, each with 64 KB of local memory. Each AIE tile is capable of computing 256 \texttt{int8} MACs per cycle, the equivalent of 58 DSP58s\cite{vek280handbook}. As a hardened ASIC, AIE can operate at up to 1 GHz, about 3.2× higher than the 312.5 MHz PL-side clock used in our setup, potentially providing a substantial performance boost. 

Achieving high performance requires careful consideration of the data movement and compute patterns.
The PL fabric supports fully custom data paths, interconnects, and fine-grained placement and routing generated by modern EDA tools, enabling data to move exactly when and where it is needed with minimal overhead. 
In contrast, AIEs operate as loosely coupled VLIW processors with fixed memory hierarchies. 
Designers are responsible for explicitly programming the workloads on each AIE tile and managing the dataflow both within a tile and across the array.
Inefficient data movement can introduce DMA stalls, memory contention, and additional buffering latency, leading to under-utilized compute cores.
These factors must be carefully managed to avoid performance degradation.

The extreme-edge scientific computing community has shown interest in using AIEs as an alternative to PL implementations, through application studies and emerging deployment toolchains~\cite{gonski2026machine,xiotidis2026amd,danopoulos2025aie4ml}. 
In particular, \aieml\cite{danopoulos2025aie4ml} demonstrates an end-to-end deployment toolchain for quantized NN, supporting spatial tiling and placement.
However, because the architectures and programming models of PL and AIEs differ fundamentally, two key questions must be addressed for AIEs to be used effectively in this domain:

\begin{itemize}
    \item \textbf{When to deploy?} PL resource congestion forces a performance-resource trade-off in \hlsml. Since AIEs follow a different design space and scaling behavior, it is not immediately clear when they should replace PL. Under what workload sizes and resource budgets does AIE deployment become more beneficial than PL?

    \item \textbf{How to deploy?} Extreme-edge scientific applications introduce distinct workload and dataflow patterns. Given the many spatial and API-level tiling and data-movement options on AIE, how should we design dataflow optimizations for this domain?

\end{itemize}

Our work studies these questions.
Section \ref{sec: micro-benchmarking} answers \emph{when} to deploy through architectural characterization and micro-benchmarking. Section \ref{sec:how} proposes dataflow optimizations to address \emph{how} to deploy extreme-edge scientific NNs.


\subsection{Experimental Setup}
All experiments are performed on an AMD-Xilinx Versal VEK280 board equipped with an AIE-ML array. AIE performance is measured using cycle-accurate hardware emulation. For a fair comparison, the PL baselines are implemented on the same VEK280 device. We extend the \hlsml framework to support Versal platforms. Performance and resource utilization are reported using AMD Xilinx Vitis and Vivado 2025.2 synthesis and implementation tools.

\section{Architectural Characterization and Micro-benchmarking}
\label{sec: micro-benchmarking}

\SetKwInOut{Input}{input}
\SetKwInOut{Output}{output}
\SetKwFor{SpatialFor}{spatial for}{:}{}
\SetKwFor{For}{for}{:}{}


This section addresses the question of \emph{when it is beneficial to use AIE}. We begin by exploring the \hlsml design space and identifying where its scalability breaks down, limiting the effective deployment of larger NNs. 
We then compare the different \hlsml and AIE design spaces and present results from micro-benchmarking experiments. 
Finally, we extract cross-domain performance and resource trends from these micro-benchmarks, which also motivates the dataflow choices elaborated in Section \ref{sec:how}.


\subsection{Reuse Factor and the PL Resource Wall}
\label{subsec:scalability}



The primary design choice for \hlsml is the user-defined \emph{reuse factor ($rf$)}, which adjusts parallelism in the layer datapaths. 
\hlsml converts ML models defined in PyTorch, TensorFlow, or Keras into HLS projects, where the $rf$ changes pipeline initiation interval (II) and loop-unrolling HLS directives. 
This guides the HLS tools in trading parallelism for resource savings and creates a large performance–resource design space. Logic and physical synthesis are handled automatically by backend EDA tools.


For small NNs, hls4ml provides high performance because the design can be fully parallelized.
As the workload grows, however, fitting the design into the available PL resources requires increasing the reuse factor and time-multiplexing arithmetic units. 
A larger reuse factor therefore trades performance for resource savings and reduces throughput.

Beyond the per-layer reuse factor, \hlsml also exposes two higher-level strategies: \emph{Latency} and \emph{Resource}. 
The Latency strategy prioritizes parallelism and relies heavily on LUTs and FFs. The Resource strategy is more conservative and makes greater use of resources such as BRAM when resources become constrained.

\begin{figure}
    \centering
    \includegraphics[width=0.89\linewidth]{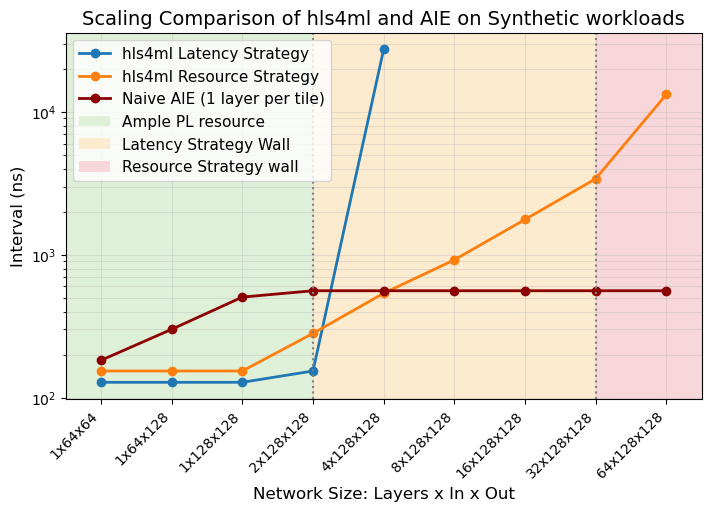}
    \caption{\hlsml performance scalability. Performance is measured by Interval, i.e., the time between output batches in steady-state execution. A smaller Interval indicates higher throughput and thus better performance. In the resource-abundant regime, \hlsml can fully parallelize the design, so Interval remains nearly constant while resource consumption increases with workload sizes. In the constrained-resource regime, arithmetic units must be time-multiplexed with a reuse factor, thereby reducing throughput and increasing the interval. AIE performance is also shown for reference.
    }
    \label{fig:LUT-wall}
\end{figure}

We construct a synthetic workload of dense layers and examine how performance scales with workload size in \hlsml (see Fig.~\ref{fig:LUT-wall}). 
As workload size increases, both strategies see an increase in the interval (the time between output batches in steady-state execution), especially when the design no longer fits on the PL with $rf{=}1$. Under the Latency strategy, the interval rises sharply, and resources are quickly exhausted. Under the Resource strategy, the interval grows steadily, making it a more practical option for medium-sized NNs that require a resource-latency trade-off.

This trend also motivates our choice of the Resource strategy as the PL baseline in later comparisons with AIE. 
Between the two \hlsml strategies, Resource is the more scalable and resource-efficient implementation, providing a more meaningful comparison. 
If PL already outperforms AIE under the Resource strategy for small workloads, then the Latency strategy would only strengthen that result. 
Conversely, once AIE outperforms the Resource strategy at larger scales, it will necessarily outperform the Latency strategy as well, since the latter reaches the PL resource wall much earlier.

We also plot an AIE implementation as a reference.
Here we use a naive mapping that deploys one layer to one AIE tile.
Since AIE resources remain abundant in this regime, the initiation interval is determined by layer size rather than number of layers.
Together, these results show that \hlsml is highly effective at small scale, but its scalability ends relatively early, motivating the use of AIE for larger workloads.

\subsection{Design Space Divergence and Micro-Benchmarking}

\begin{algorithm}[t]
\caption{\emph{LARE} metric for dense layers $(n_\mathrm{in}, n_\mathrm{out})$}
\label{alg:LARE}
\Input{Dense layer shape: input $n_\mathrm{in}$, output $n_\mathrm{out}$}

\For{$rf \in \mathrm{Legal\ reuse\ factors\ for\ } n_\mathrm{in}, n_\mathrm{out}$}{
    $\mathrm{Find\ PL\ resource\ consumption\ } R_{\mathrm{PL}}(n_\mathrm{in}, n_\mathrm{out}, rf)$

    $\mathrm{Find\ PL\ performance\ } P_{\mathrm{PL}}(n_\mathrm{in}, n_\mathrm{out}, rf)$
}

$\mathrm{Find\ AIE\ performance\ } P_\mathrm{AIE}(n_\mathrm{in}, n_\mathrm{out})$

$\textbf{Interpolate}\ \mathrm{and\ find}$

$\quad rf_{\mathrm{eq}}\ \mathrm{s.t.}\ P_\mathrm{AIE}(n_\mathrm{in}, n_\mathrm{out})
= P_{\mathrm{PL}}(n_\mathrm{in}, n_\mathrm{out}, rf_{\mathrm{eq}})$

\Output{$\emph{LARE} = R_{\mathrm{PL}}(n_\mathrm{in}, n_\mathrm{out}, rf_{\mathrm{eq}})$}
\end{algorithm}

\begin{figure}
    \centering
    \includegraphics[width=0.89\linewidth]{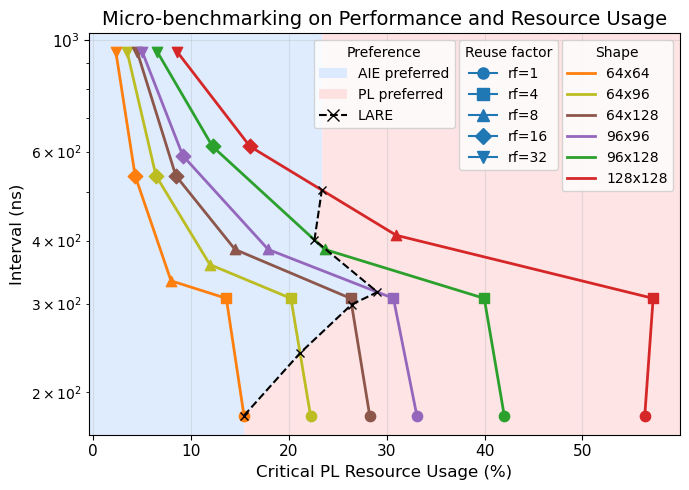}
    \caption{Micro-benchmarking to understand resource–latency trade-off. Each colored line represents one layer shape on PL; each point on a colored line corresponds to a different reuse factor. Larger reuse factor (top-left) results in less resource usage and higher interval (weaker performance). The black points on colored lines mark the AIE performance ($y$ axis) and LARE value ($x$ axis) of the corresponding layer. The blue region represents congested PL resource, where large reuse factor is required for PL deployment, and the AIE achieves better latency. The red region indicates resource-redundant PL, where spatial dataflow PL implementations provide better latency than AIE.}
    \label{fig:bigbg}
\end{figure}

It is challenging to make direct comparisons between the PL and AIE design spaces. 
Section~\ref{subsec:scalability} describes how the PL performance-resource trade-off in \hlsml is primarily controlled by the per-layer reuse factor. 
AIEs follow a vector-processor programming model with a deterministic network-on-chip, and their performance depends mainly on workload dimensions together with spatial and API-level tiling. 
A meaningful comparison must address the additional PL design parameter of reuse factor, which is not present in AIEs.

To compare the two domains fairly and extract generalizable insights, we perform micro-benchmarking experiments on the atomic workload of a single dense layer mapped to one AIE tile.
Micro-benchmarking focuses on isolating layer-level design trends that translate to larger designs.
Because the reuse factor is assigned per layer, evaluating the same layer at different reuse factors effectively represents that layer under different NN resource budgets. 
As the NN size grows, the budget available to each layer shrinks, forcing higher reuse. 
In this sense, a single-layer resource and performance trade-off curve captures how that layer would behave across NNs of different sizes.

For the AIE micro-benchmark, performance is fixed for a given layer shape if the workload is mapped to one tile. 
This allows us to compare the AIE performance with the PL resource-performance trade-off curve. 
We define the minimum PL resource required for \hlsml to match the AIE performance as the \emph{latency-adjusted resource equivalent} (LARE), as shown in Algorithm \ref{alg:LARE}.

We repeat the experiment of reuse factor sweep and determining LARE for different dense layer shapes (Fig.~\ref{fig:bigbg}). For each shape, the colored curve traces the PL resource-latency trade-off, and the black point marks the PL resource required to match the AIE interval. In this sense, LARE identifies the crossover between the PL-favorable and AIE-favorable regimes for that layer shape. This metric provides:
\begin{itemize}
    \item \textbf{A decision boundary}. If the available PL resource for the layer exceeds its LARE value, then the PL implementation can match or outperform the AIE implementation. Otherwise, the resource-constrained PL will have worse performance due to inefficient reuse.
    \item \textbf{An efficiency indicator}. LARE also reflects how effectively the AIE implementation uses its compute tile resources. A low LARE means that the same performance can be matched with only a small PL budget, suggesting that the implementation is not very efficient and could benefit from further AIE optimizations. 
\end{itemize}

\begin{figure*}[!ht]
    \centering
    \includegraphics[width=0.96\linewidth]{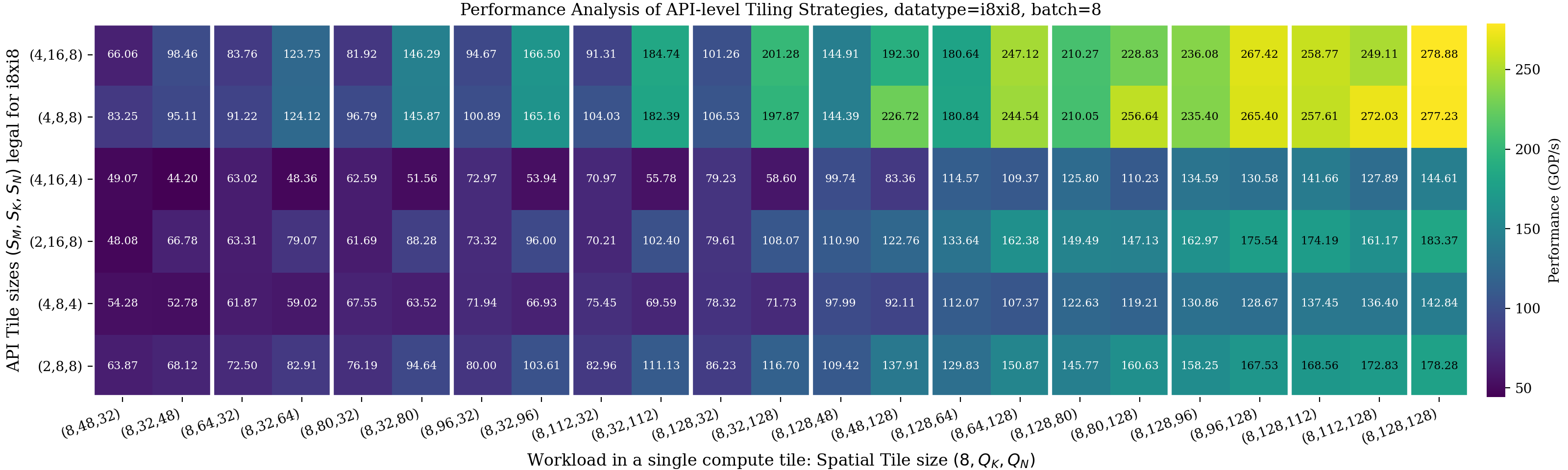}
    \caption{
    Performance analysis of GEMM workloads with batch size of 8 implemented in a single compute tile, to measure the performance impact of API-level tiling strategies. 
    Each two-column group along the x-axis corresponds to a workload size, with the number of operations increasing linearly by 4096 across groups; the two workloads within a group have the same number of operations but differ in asymmetry ($Q_K$-larger vs $Q_N$-larger). 
    The y-axis shows the \texttt{aie::mmul} API sizes $(S_M,S_K,S_N)$ that are legal for the datatype \texttt{i8$\times$i8}. 
    We observe that the APIs $(4,8,8)$ and $(4,16,8)$ achieve the best performance across all workload sizes, and spatial tiling with larger $Q_N$ consistently yields better performance.}
    \label{fig:lat_hm_api_tiling}
\end{figure*}

\subsection{Summary and Discussion}
\hlsml is effective when the model size is small and sufficient PL resources are available. 
When \hlsml becomes highly resource-constrained, the performance suffers, making AIE a preferable option. 
The PL-AIE crossover is not only determined by model size, but also by the layer shape and the resource budget available to that layer.
Micro-benchmarking experiments and the LARE metric provide a practical decision boundary for choosing between PL and AIE.

Fig.~\ref{fig:bigbg} also shows that the LARE trend is not linear, or even monotonic. 
This indicates that AIE utilization depends strongly on workload size and shape.
In this section, we consider only the naive mapping of one layer to one AIE tile, regardless of shape. 
The results, therefore, also motivate the next section: how to improve AIE performance through tiling and dataflow optimization. 

\section{AIE Tiling and Dataflow Optimizations}
\label{sec:how}

Implementing low-latency NNs on the AIE involves navigating a complex design space of architectural constraints and tiling parameters with varying trade-offs. 
The under-utilization of naive mappings observed in Section~\ref{sec: micro-benchmarking} motivates exploring strategies to distribute workloads across more AIE compute tiles to increase performance by optimally leveraging parallelism.
This section details the implementation strategy for GEMM, the primary compute workload in scientific NNs, using the \aieml framework~\cite{danopoulos2025aie4ml}. 

To satisfy microsecond-level latency constraints, we follow the established convention for extreme-edge scientific inference by using the smallest possible batch size (8 for the \texttt{int8} data type in AIE) and keeping all model weights stationary and fully on-chip.
This approach diverges from traditional GEMM implementations for larger NNs, which rely on high-throughput batching and external memory access for weights, since those workloads do not face the same strict performance constraints.
We focus on two levels of tiling: spatial tiling across compute tiles to parallelize each GEMM workload, and API-level tiling within each compute tile to efficiently utilize vector processing units.
We characterize the architectural constraints of AIE and propose design rules for API-level and spatial tiling based on empirical benchmarking.
We also quantify the latency cost of crossing the PL-AIE boundary and present a design rule for hybrid implementations.

\subsection{Two-Level GEMM Tiling}
\label{sec:gemm}

The standard implementation of an $M,K,N$ GEMM workload $A_{(M,K)}{\times}B_{(K,N)}{=}C_{(M,N)}$ in AI Engines follows the two-level tiling hierarchy detailed in Algorithm~\ref{alg:two_level_gemm}~\cite{danopoulos2025aie4ml}. 

\subsubsection{Spatial Level}
The global workload is partitioned across $P_K{\times}P_N$ compute tiles of the AIE array. 
The $K$ and $N$ dimensions are divided into $P_K$ and $P_N$ compute-tile columns and rows, respectively.
Each compute tile gets a spatial tile workload of size $M{\times}Q_K{\times}Q_N$.

\subsubsection{API Level}
In each compute tile, the workload is further subdivided at the API level into smaller blocks of size $(S_M, S_K, S_N)$. 
This tuple must be one of the few legal tuples supported by the \verb|aie::mmul| API. 
The API is called $R_M, R_K, R_N$ times over time, to process the spatial tile. 

This formulation defines the global workload dimensions to physical hardware mapping, serving as the basis for the performance analysis and design rules in the following sections.

\SetKwInOut{Input}{input}
\SetKwInOut{Output}{output}
\SetKwFor{SpatialFor}{spatial for\_2d}{:}{}
\SetKwFor{ForD}{for\_2d}{:}{}
\SetKwFor{For}{for}{:}{}

\begin{algorithm}[t]
\caption{Tiling GEMM in two levels. Spatial (tiles, size):$(P,Q)_{K,N}$ \& API (tiles, size):$(R,S)_{M,K,N}$}
\label{alg:two_level_gemm}
\Input{$A \in \mathbb{R}^{M \times K}$, $B \in \mathbb{R}^{K \times N}$ -- Input matrices}
\Input{$(P_K, P_N)$ -- No. of compute tiles (col, row)}
\Input{$(S_M, S_K, S_N)$ -- Legal API tile sizes}
\Output{$C \in \mathbb{R}^{M \times N}$ -- Output matrix}

$(Q_K, Q_N) = (K / P_K,\; N / P_N)$\;

$(R_M, R_K, R_N) = (M / S_M,\; Q_K / S_K,\; Q_N / S_N)$\;

$C = 0$\;

\SpatialFor{$(p_k, p_n) \in [0, P_K) \times [0, P_N)$}{
    $(k_q, n_q) = (p_k Q_K,\; p_n Q_N)$\;
    
    \ForD{$(r_m, r_n) \in [0, R_M) \times [0, R_N)$}{
        \tcp{2x unrolled in implementation}
        $(m, n) = (r_m S_M,\; n_q + r_n S_N)$\;
        
        $C_t = 0$\;
        
        \For{$r_k \in [0, R_K)$}{
            $k = k_q + r_k S_K$\;
            
            $A_t = A[m{:}m{+}S_M,\; k{:}k{+}S_K]$\;
            
            $B_t = B[k{:}k{+}S_K,\; n{:}n{+}S_N]$\;
            
            $C_t \mathrel{+}= \texttt{aie::mmul.mac}(A_t, B_t)$\;
        }
        $C[m{:}m{+}S_M,\; n{:}n{+}S_N] \mathrel{+}= C_t;$\tcp*[f]{cascade}
     \vspace{-0.35em}
    }
}
\end{algorithm}

\subsection{Architectural Constraints}

Extreme-edge scientific workloads typically receive data directly into the PL~\cite{qubit,tokamak}.
To process this data in AI Engines, it must be streamed through the Programmable Logic I/O (PLIO) interfaces. 
At a typical clock frequency of 312.5 MHz, a 128-bit PLIO provides a bandwidth of 5 GB/s.

The AIE array of VEK280 consists of 38 columns and 8 rows of compute tiles. 
The current version of the \aieml toolchain restricts usage to 31 columns (indices 7 to 37) to minimize routing effort, since PLIOs are placed starting there. 
Within the AIE array, each compute tile contains a scalar processor and a vector unit capable of 256  $\texttt{int8}{\times}\texttt{int8}$ MACs per cycle optimized for fixed-point arithmetic. 
AMD provides high-level API calls to program vector units for matrix multiplication: \texttt{aie::mmul<m,k,n,dtype,dtype>}~\cite{aie_api,UG1079}.

For data movement, every compute tile has (1) one input and one output port for streaming, each 32-bit; (2) a 512-bit cascade bus that allows partial sums to be moved from west to east for accumulation; and (3) a 64 KB local data memory accessible via two 256-bit load units and one 256-bit store unit per cycle at 1 GHz, providing a local read and write bandwidth of 64 GB/s and 32 GB/s, respectively.

\subsection{API-Level Tiling Optimization}

With hardware constraints in mind, we first study GEMM decomposition within a single compute tile to determine design rules for the optimal tile size $(S)$ and count $(R)$ in API-level tiling.
To maximize vector-unit utilization, the GEMM inside a compute tile is manually unrolled $2 {\times} 2 {\times} 2$ times, making the effective tile size twice the base API tile sizes.

Fig.~\ref{fig:lat_hm_api_tiling} shows the results of our benchmarking experiment, which sweeps the size and shape of a workload $(8,Q_K,Q_N)$ in a single compute tile and measures the performance in GOP/s for all legal API sizes.
We find that $(4, 8, 8)$ and $(4, 16, 8)$ consistently outperform other API tile shapes. 
While their performance is similar, $(4, 8, 8)$ offers finer granularity, as it permits a minimum $K$ of 16 (due to $2{\times}$ unrolling), compared to 32 for the $(4, 16, 8)$ shape. Therefore, we choose $S{=}(4, 8, 8)$ for further analysis.
We summarize this design rule as:

\vspace{-1em}
\begin{heuristic}
For extreme-edge scientific computing, a default API tile size of $S=(4,8,8)$ offers the best overall performance and granularity.
\end{heuristic}

Furthermore, Fig.~\ref{fig:lat_hm_api_tiling} also demonstrates the impact of workload asymmetry. 
Performance is up to 2$\times$ higher when $Q_N > Q_K$ than when $Q_K > Q_N$ for a fixed single-compute-tile workload size, and this advantage carries over to the global workload.
This is because the wide vector processing units and output accumulators are better utilized when the output dimension is larger.

\vspace{-1em}
\begin{heuristic}
Compute-tile performance for a fixed workload size (MAC count) varies with workload shape. 
API tiling should prioritize the $N$ dimension (output channels) over the $K$ dimension (input reduction).
\end{heuristic}

\subsection{Spatial Tiling Optimization}
\label{sec:spatial_tiling}

Having identified an efficient API shape, we next examine how the workload should be mapped across the AIE array, shifting the analysis to array-level parallelism and its implications for performance.
Spatial tiling partitions the global workload into $Q_{K,N}$-sized sub-workloads and assigns them to $P_{K,N}$ columns and rows of compute tiles, as defined in Algorithm \ref{alg:two_level_gemm}. 
The dedicated cascade bus propagates partial sums between compute tiles, and the sub-workloads are assigned west to east to minimize data movement.

\begin{figure}[!t]
    \centering
    \includegraphics[width=.9\linewidth]{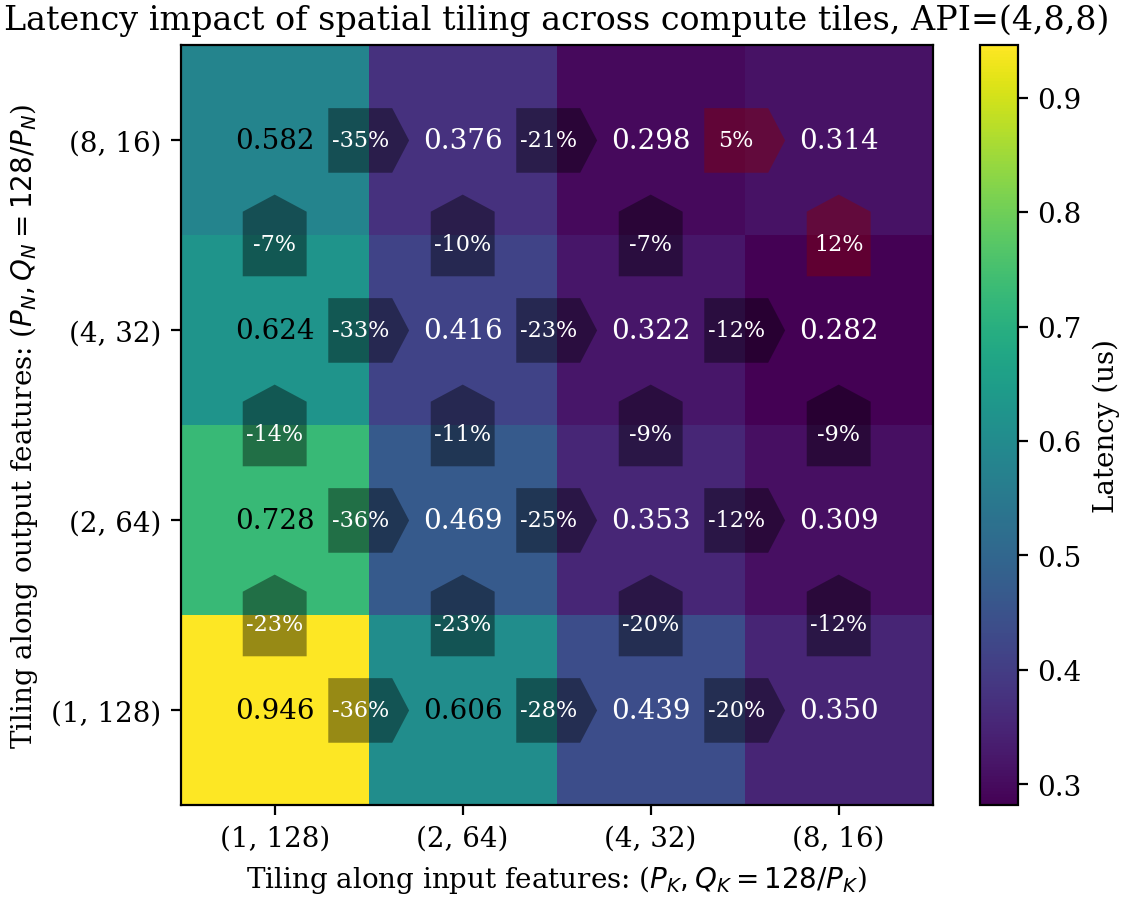}
    \caption{Latency reduction of tiling a GEMM workload / dense layer of size (8,128,128) spatially, across $P_K$ columns and $P_N$ rows of compute tiles. The API tile size $(S_M,S_K,S_N)$ is fixed to (4,8,8). Each arrow shows the percentage change in latency upon doubling the number of compute tiles along row/column in the AIE array.}
    \label{fig:lat_hm_spatial_tiling}
\end{figure}

Fig.~\ref{fig:lat_hm_spatial_tiling} shows the benchmarking results of spatially tiling a workload of global size $(8,128,128)$ across varying numbers of columns and rows of compute tiles, with API size fixed to $(4,8,8)$.
Along the \emph{negative-slope diagonals} $(i+j{=}\mathrm{const})$, the factor of parallelism, i.e., the number of compute tiles per layer, ($P_K{\times}P_N)$ stays constant.
We observe that for a fixed factor of parallelism $P_K P_N$, configurations that allocate more compute tiles along the $K$ dimension consistently achieve lower latency than those that allocate more compute tiles along the $N$ dimension. 
This matches the trend at the API-level favoring smaller $R_K$. 

\vspace{-1em}
\begin{heuristic}
    Spatial tiling a given layer across the AIE array should prioritize column-wise expansion, with row-wise tiling used more selectively.
\end{heuristic}

Next, we observe that while more parallelism generally yields higher performance (as expected), performance scales much more slowly with the number of compute tiles.
Each arrow shows the effect of doubling the number of compute tiles along either row or column direction of the AIE array.

\vspace{-1em}
\begin{heuristic}
    Spatial tiling exhibits diminishing returns. A workload of $(M,Q_K,Q_N){=}(8{\times}32{\times}64)$ per compute tile provides reasonably strong performance; further spatial tiling yields ${<}15\%$ gains while requiring twice the tiles.
\end{heuristic}

Although increasing spatial parallelism initially increases performance, the improvement is not monotonic. 
In the measured design space, performance is maximized at a spatial tile size of $(Q_K,Q_N)=(16,32)$, whereas further spatial tiling reduces performance.

\vspace{-1em}
\begin{heuristic}
    Using more compute tiles does not necessarily improve performance. To avoid under-utilization, the per-tile workload should be at minimum $8{\times}16{\times}32$.
\end{heuristic}

\subsection{The Cost of Column Exhaustion}

\begin{figure}[!t]
    \centering
    \includegraphics[width=0.4\linewidth]{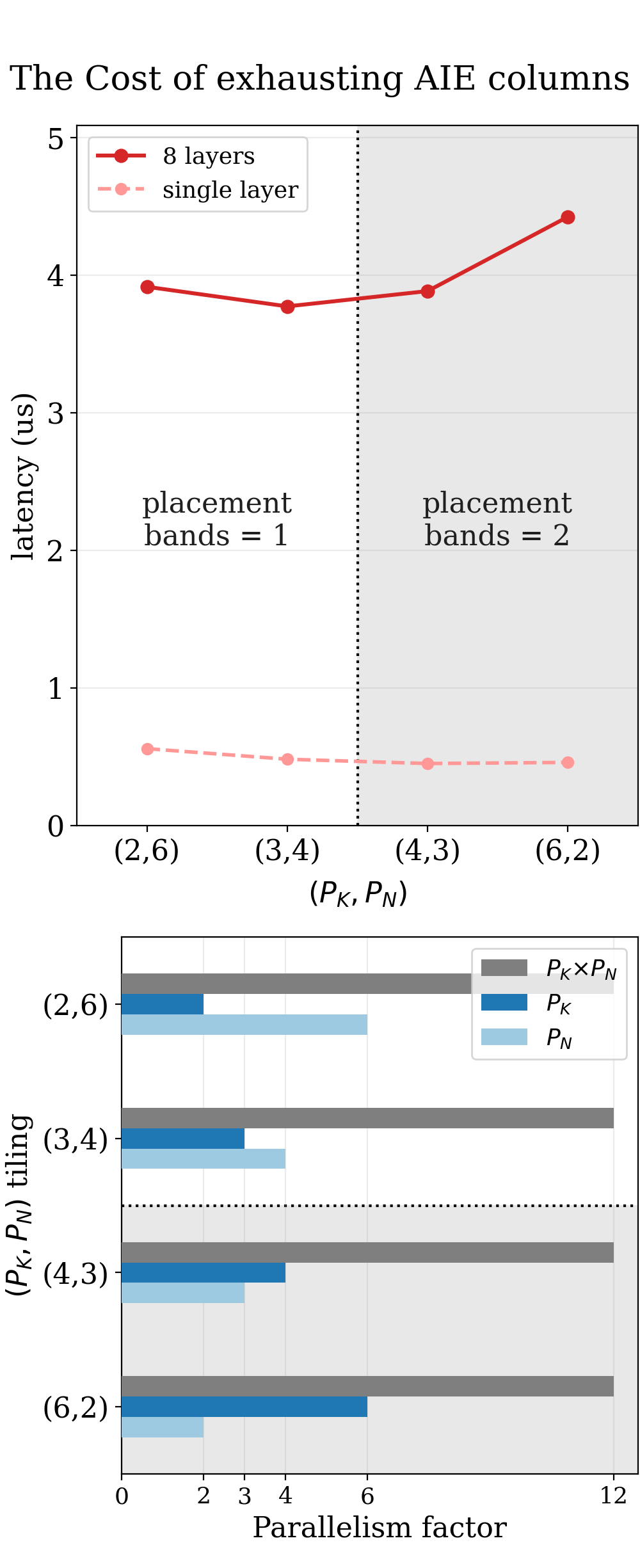}
    \includegraphics[width=0.58\linewidth]{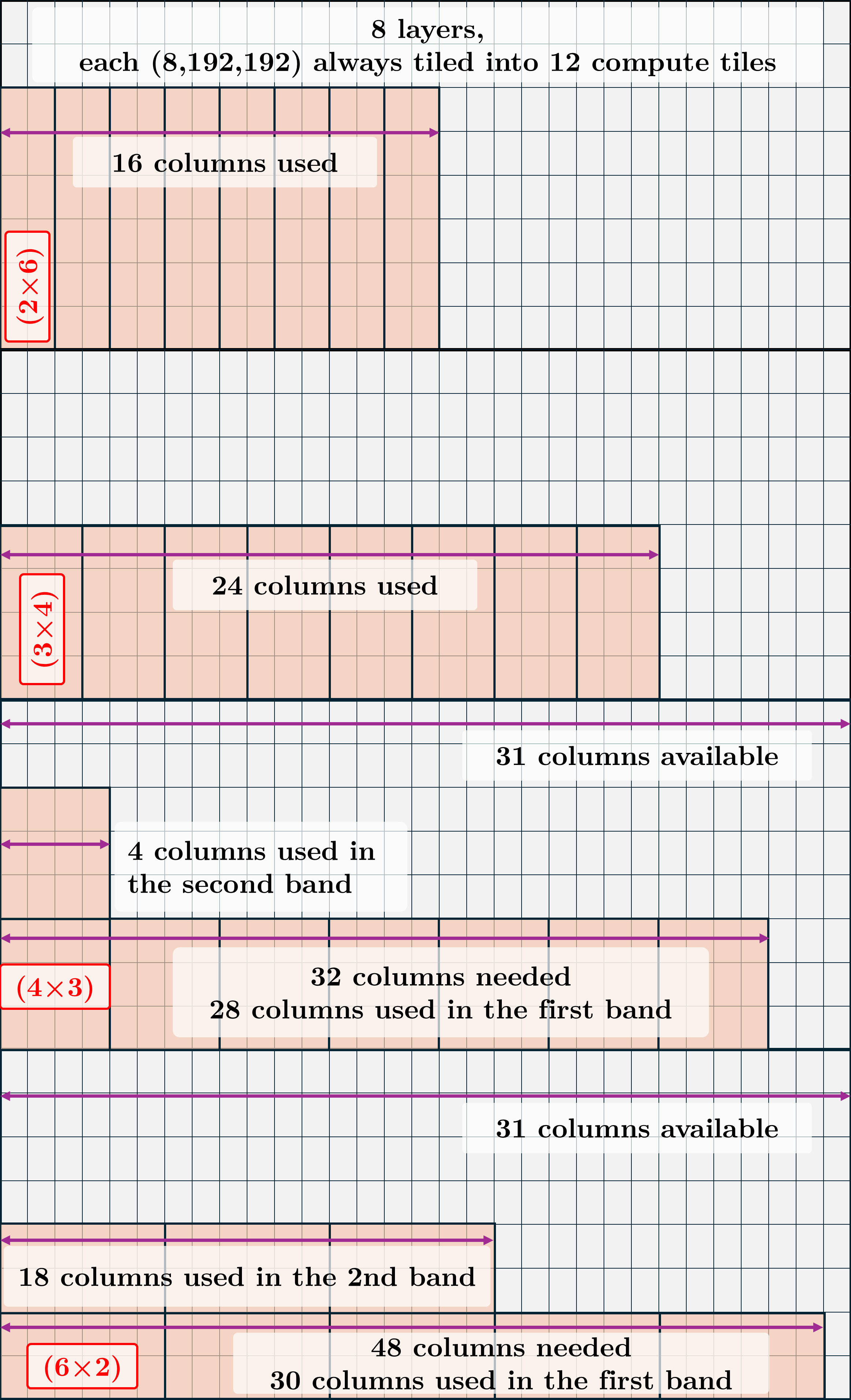}
    \caption{
    Latency impact of exhausting the available AIE columns. 
    We use a dense model with 8 layers, each of size (8,192,192), datatype \texttt{i8$\times$i8}, and API tile size $(4,8,8)$ each implemented in $P_K{\times}P_N{=}12$ compute tiles so other latency factors remain unchanged.
    We then vary the spatial tiling asymmetry $(P_K,P_N)$.
    The results show that column-wise expansion improves performance only up to the single-band column limit; once layers spill into multiple bands, resource contention introduces a latency penalty.}
    \label{fig:asym_sweep}
\end{figure}

The previous results suggest that column-wise expansion is beneficial, but that trend cannot continue indefinitely on a finite array.
We therefore examine the performance implications of this preferred spatial tiling strategy once it exceeds the physical width of the device.
When this occurs, subsequent layers must be placed in ``bands''~\cite{danopoulos2025aie4ml}, which are additional rows of compute tiles above the initial layers (Fig.~\ref{fig:asym_sweep}). 
This architectural shift results in a measurable performance loss.

To quantify the impact of column exhaustion, we conduct a sweep of spatial-tiling asymmetries in Fig.~\ref{fig:asym_sweep}. 
We used a constant model with 8 layers, 192 input/output features, and a batch size of 8. 
This configuration was selected to maximize the number of available sweep points while remaining within AIE hardware constraints. 
The datatype was fixed at $\texttt{i8}{\times}\texttt{i8}$ using the $(4, 8, 8)$ API. 
To account for manual unrolling by a factor of $2\times$ and prevent latency impact from uneven spatial tiling, we ensured that each compute tile processed a minimum workload of $(8, 16, 16)$.
The current implementation of \aieml limits the usable array width to 31 columns. 
Nevertheless, the analysis is not tied to this specific limit and would remain valid if this restriction were relaxed, with the relevant transition occurring at a different column count.

In this experiment, the total parallelization factor per layer was fixed at $P_KP_N{=}12$, while varying the spatial-tiling asymmetry $(P_K,P_N){=}$(cols, rows) of compute tiles per layer:

\begin{itemize}
    \item $P_K{\times}P_N{=}2{\times}6$: This requires 16 total columns ($8 \text{ layers} \times 2 \text{ columns}$). 
    Since this fits within the 31-column limit, all layers are placed in a single band.
    \item $P_K{\times}P_N{=}3{\times}4$: This requires 24 columns. 
    It still fits within a single band and demonstrates improved performance over the 2-column configuration, following the trend established in Section ~\ref{sec:spatial_tiling}.
    \item $P_K{\times}P_N{=}4{\times}3$: This requires 32 columns, exceeding the 31-column limit.
    Consequently, \aieml places 7 layers in the first band and the 8th layer in a second band above the first. 
    This forces layers across different bands to share the same set of memory tiles, leading to resource contention and performance degradation.
    \item $P_K{\times}P_N{=}6{\times}2$: This requires 48 columns, where 5 layers are placed in the first band and 3 in the second, further worsening the latency.
\end{itemize}

These results demonstrate that, while column-wise spatial tiling is generally superior, its benefits are lost once the model exceeds the array's physical width. 

\vspace{-1em}
\begin{heuristic}
    Column exhaustion is costly. A model implementation should spatially tile each layer across as many AIE columns as possible (maximizing $P_K$), while not exceeding the column limit and utilizing only one band.
\end{heuristic}

\subsection{The Cost of Crossing the Fabric Boundary}

\begin{figure}[!t]
    \centering
    \includegraphics[width=1\linewidth]{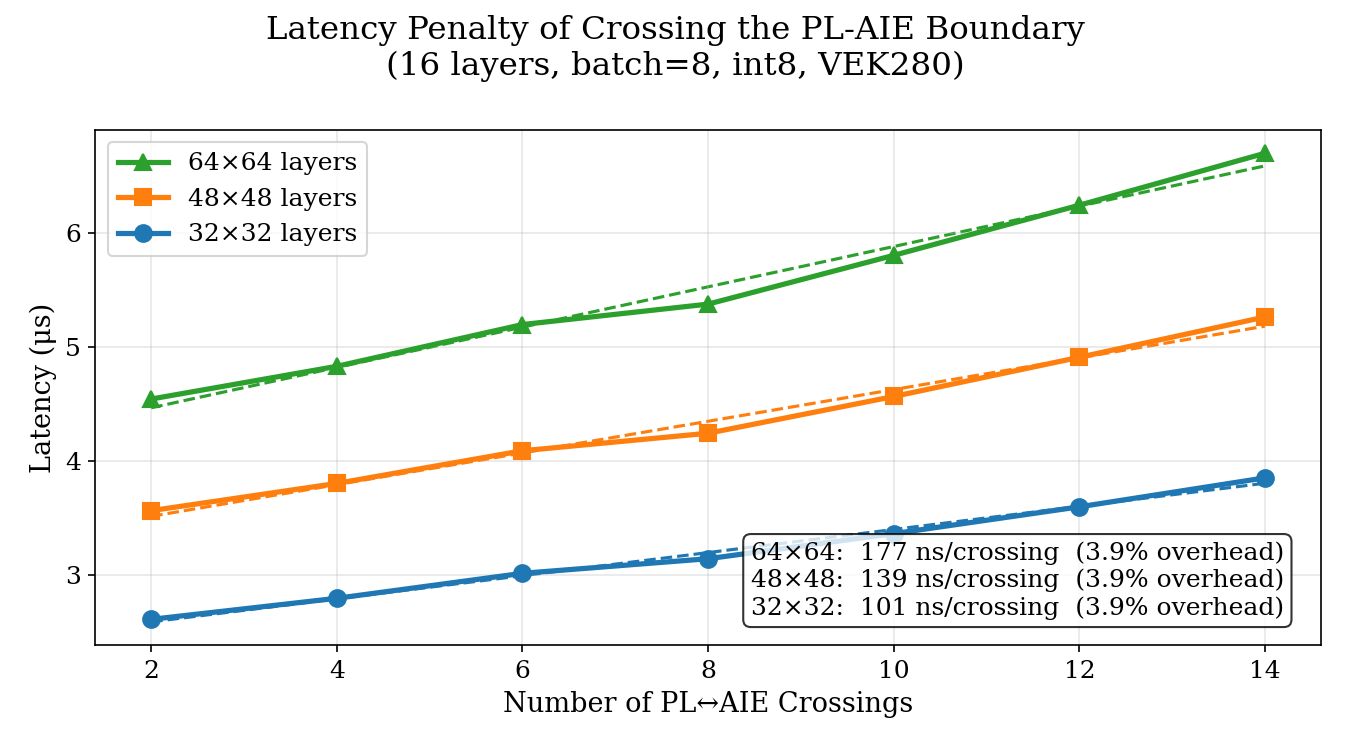}
    \caption{Latency overhead of crossing the AIE-PL boundary. Each experiment has the same dense model with 16 equal layers, always 8 layers in AIE and 8 layers in the PL, varying only the number of crossings. The gradient of the line fit to total latency suggests a latency overhead of 3.9\% per crossing. }
    \label{fig:line_lat_cross}
\end{figure}

Versal devices integrate a PL and an AIE array on the same die, making heterogeneous implementations a design option. 
The AIE array is particularly well suited to regular, vectorizable dense workloads such as GEMM, which align with its vector-oriented programming model and matrix-multiplication support~\cite{UG1079,maxeva,danopoulos2025aie4ml}. 
By contrast, the PL remains a flexible domain for implementing custom datapaths and auxiliary functions such as bit-manipulating logic and fast data reordering using RTL or HLS~\cite{amd_ug1504_2025,amd_ug1273_designguide_2025,amd_ug1399_hls_2025}.
Recent AIE--PL hybrid designs have also used the PL to implement nonlinear activation functions~\cite{sapkas2026gru}. 

A designer may, therefore, place nonlinear or other auxiliary stages in the PL while keeping GEMM-dominated stages in the AIE. 
For such a heterogeneous implementation, a fundamental question is the latency penalty of crossing the PL--AIE boundary.
We therefore quantify this overhead directly. 

For this, we constructed a dense model with 16 layers, $\texttt{i8}{\times}\texttt{i8}$ datatype, and batch size 8, with input/output dimensions fixed per sweep. 
To keep the computational cost constant across all experiments, each configuration used exactly 8 layers in the PL and 8 layers in the AIE, with the first and last layers always placed in the PL. 
Because current versions of \hlsml and \aieml do not natively support automated model partitioning, we manually extracted the dense implementations from both flows and composed a hybrid pipeline~\cite{hls4ml2021,danopoulos2025aie4ml}. 
AIE layers were mapped one per compute tile, while PL layers used a reuse factor of 1. 
We then swept the number of PL-AIE boundary crossings from 2 to 14 with a stride of 2. 
The total latency was obtained by combining the PL latency reported by HLS synthesis with the AIE latency reported by cycle-accurate AIE simulation.

The results in Fig.~\ref{fig:line_lat_cross} show a strong linear trend with $R^2{=}0.98$, indicating that our experiment isolates the latency penalty of crossing the boundary. 
For every workload, each additional crossing contributes a latency overhead of approximately 3.9\% relative to the baseline case of two crossings (input and output only). 
This result provides a quantitative penalty term that any designer considering a PL-AIE split should incorporate into the mapping decision.

\vspace{-1em}
\begin{heuristic}
For hybrid PL-AIE pipelines, each additional PL-AIE boundary crossing adds about 3.9\% latency relative to the two-crossing baseline. 
A heterogeneous split is worthwhile only when the computational benefit of placing a stage in its preferred domain exceeds this cost.
\end{heuristic}

\section{Full NN Deployment}
\label{sec:nn}

We seek to answer two practical questions on real extreme-edge scientific workloads: (1) Can AIE serve as a competitive alternative once PL implementations are forced into high reuse by resource congestion? (2) How much performance gain can be achieved through the design rules developed in Section~\ref{sec:how}?

We evaluate three workloads whose PL implementation cannot satisfy the performance goals shown in Fig.\ref{fig:intro}: the VAE at LHC~\cite{vaelarge}, the multi-Qubit readout discriminator~\cite{qubit}, and the large autoencoder from the MLPerf Tiny benchmarks~\cite{banbury2021mlperf}.
All models use 8-bit quantization, and the results are summarized in Table~\ref{table:aie-compute-capability}, with performance reported in millions of inferences per second (MHz). 
We observe that the naive AIE implementation (one compute tile per layer) is competitive with PL designs that require high reuse, while applying our design rules yields up to a 4$\times$ performance improvement. 
While the PL implementation cannot meet the 40 MHz extreme-edge collision rate target~\cite{cms2016cms}, the optimized AIE designs exceed it.


\begin{table}[!t]
\centering
\begin{tabular}{l c c c c c  }
\hline
\textbf{Metric} & \textbf{VAE}\cite{vaelarge} & \textbf{Qubit}\cite{qubit} & \textbf{Autoencoder}\cite{banbury2021mlperf} \\

\hline
\textbf{MACs}               & 34.8k & 82.9k & 116.7k \\
\textbf{Min. \hlsml r.f.}        & 8     & 16    & 32 \\
\textbf{PL Perf. (MHz)}   & 20.8   & 12.5   & 8.4  \\
\textbf{Naive AIE Perf. (MHz)}   & 22.7 & 14.4            & 15.9   \\
\textbf{Optimized Perf. (MHz)} & 97.9 & 58.9           & 58.8   \\
\hline
\end{tabular}
\vspace{2 pt}
\caption{Full NN Deployment Results. }
\label{table:aie-compute-capability}
\vspace{-3em}
\end{table}

\section{Related Work}
\label{sec:related}



Several recent works have investigated the use of AIEs and AMD Versal ACAP platforms for large ML workloads. 
CHARM\cite{charm, charm2.0} proposes analytical models and conducts design space exploration to balance workloads in different-sized accelerators to support variable-size matrix multiplications in transformer layers. 
MaxEva~\cite{maxeva} explores hard-wiring datapaths between AIE tiles, allowing higher performance at the cost of increased resource consumption and reduced flexibility. 
XVDPU~\cite{xvdpu} proposes a methodology aimed at high performance by leveraging AIE tiles to compute convolutions while using the PL for auxiliary functions like buffering, data movements, and scheduling. 
SPARTA~\cite{sparta} leverages the Multi-Level Intermediate Representation (MLIR) compiler framework to support the development of spatial accelerators for horizontal diffusion weather stencil prediction. 
Brown~\cite{brown2023exploring} explored optimal approaches for structuring AIE kernels in Versal platforms and their interface with the PL to accelerate atmospheric modeling, showing increased performance compared with traditional AMD Alveo U280 platforms. 
Noah et al.~\cite{space-aie} evaluated the use of AIEs for space applications, mainly targeting CNNs and FFT, demonstrating their effectiveness in providing speedup and power efficiency.
Chen et al.~\cite{chen2023exploiting} leveraged the heterogeneity of Versal platforms to optimize AI inference for Graph Neural Networks, proposing a custom hardware module for sparse primitives to be deployed in PL, combined with efficient computation of the dense primitive on AIEs. 
Yang et al.~\cite{yang2023aim} proposed analytical models and code generation to automate and optimize the mapping of arbitrary-precision integer multiplication in Versal ACAP platforms leveraging PL, AIEs, and CPU.
Zhang et al.~\cite{zhang2024cat} presented a customized abstract transformer accelerator family for Versal platforms, which can be efficiently mapped into a customized accelerator for a target transformer architecture.

Prior work has predominantly focused on optimizing AIEs for large-scale matrix multiplications. To the best of our knowledge, this paper is the first to investigate the design rules of AIEs under the strict low-latency and high-performance requirements of extreme-edge scientific computing workloads.

\section{Conclusion}
\label{sec:conclusion}
This work addresses \emph{when} and \emph{how} NNs for extreme-edge scientific computing should be implemented on AIEs.
While the implementation of small networks on PL using \hlsml remains effective, resource scaling for larger networks forces higher reuse and worse performance.
We introduce the LARE metric that quantifies the regime where the deployment on AIEs becomes the better option.
We then derive design rules for spatial and API-level tiling, and analyze the performance impacts of workload asymmetry, diminishing returns from over-tiling, AIE column exhaustion, and PL-AIE boundary crossings.
Finally, as summarized in Fig.~\ref{fig:intro} and Table~\ref{table:aie-compute-capability}, our design rules enable larger real-world models that were previously unable to meet the strict extreme-edge performance targets on PL to exceed those targets on AIEs, thereby expanding the frontier of fully on-chip inference.

\clearpage
\newpage
\bibliographystyle{IEEEtran}
\bibliography{references}

\end{document}